\documentstyle[aps,prl,multicol]{revtex}
\begin{document}
\draft
\title{Exact results for a tunnel-coupled pair of 
trapped Bose-Einstein condensates}

\author { 
Huan-Qiang Zhou\cite{email0}, Jon Links,  Ross
H. McKenzie, and Xi-Wen Guan${\dagger}$}

\address{Centre for Mathematical Physics, The University of Queensland,
		     4072, Australia \\
		     $\dagger$Instituto de F\'{\i}sica da UFRGS, Av. Bento
		     Gon\c{c}alves 9500, Porto Alegre, RS - Brazil}

\maketitle

\vspace{10pt}

\begin{abstract}
A model describing coherent quantum tunneling between two trapped Bose-Einstein
condensates is shown to admit an exact
solution. 
The spectrum is obtained by the
algebraic Bethe ansatz. An asymptotic analysis of the Bethe ansatz
equations leads us to explicit expressions for the energies of the 
ground and
first excited states in the limit of {\it weak} tunneling and all
energies for {\it strong}
tunneling. The results are used to extract the asymptotic limits of the
quantum fluctuations of the boson
 number difference between the two Bose-Einstein
condensates and to  characterize the degree of
coherence in the system. 
\end{abstract}




\def\a{\alpha}
\def\b{\beta}
\def\d{\dagger}
\def\e{\epsilon}
\def\g{\gamma}
\def\K{\kappa}
\def\ap{\approx}
\def\l{\lambda}
\def\o{\omega}
\def\t{\tilde{\tau}}
\def\s{S}
\def\D{\Delta}
\def\L{\Lambda}
\def\T{{\cal T}}
\def\TT{{\tilde{\cal T}}}

\def\beq{\begin{equation}}
\def\eeq{\end{equation}}
\def\bea{\begin{eqnarray}}
\def\eea{\end{eqnarray}}
\def\ba{\begin{array}}
\def\ea{\end{array}}
\def\no{\nonumber}
\def\le{\langle}
\def\re{\rangle}
\def\lt{\left}
\def\rt{\right}

\newcommand{\reff}[1]{eq.~(\ref{#1})}

\begin{multicols}{2}

Macroscopic quantum tunneling 
is one of the most fascinating phenomena in condensed matter
physics \cite{fpctl,science}.
Experimental realization of  Bose-Einstein condensates (BEC's) in
dilute atomic alkali gases has stimulated 
a diverse range of theoretical and experimental  
research activity \cite{pw98,dgps99,l01,otfyk,cbfmmtsi}. 
A particularly exciting possibility is that
a pair of BEC's (such as a BEC trapped
in a
double-well potential) may provide a model tunable system
in which to observe macroscopic quantum tunneling.
 In this Letter we show that the canonical
Hamiltonian for a pair of
 tunnel-coupled BEC's \cite{l01} has an exact solution.
The model is also realizable in  Josephson-coupled 
super-conducting metallic nano-particles \cite{schon}. 
This connects the model to the rich and powerful
mathematics associated with the algebraic Bethe ansatz \cite{kib}
and provides a means to study the degradation of 
the  coherence between the two BEC's
 which occurs as the
tunnel coupling increases (or the size of the BEC
decreases). The exactness of our approach means
that it  is not necessary to resort to approximations such as
Gross-Pitaevskii mean-field theory and the phase-number
formulation which become questionable
in the coherent-incoherent crossover region.

The canonical Hamiltonian takes the form \cite{l01}  
\bea
H&=& \frac {K}{8}  (N_1- N_2)^2 - \frac {\Delta \mu}{2} (N_1 -N_2)\no\\
&& -\frac {\cal {E} _J}{2} (a_1^\dagger a_2 + a_2^\dagger a_1).
\label {ham} \eea
where $a_1^\dagger, a_2^\dagger$ denote the single-particle creation
operators in the two wells and  $N_1 = a_1^\dagger a_1, 
N_2 = a_2^\dagger a_2$ are the corresponding
boson number operators. The total boson number $N_1+N_2$
is conserved and set to the fixed value of $N$. 
The physical meaning of the coupling parameters
for different realizable systems 
may be found in Ref. 
\cite{l01}. 
It is useful to divide the parameter
space into three regimes; viz. Rabi ($K/{\cal E}_J<< N^{-1}$),
Josephson ($N^{-1}<<K/{\cal E}_J<<N$) and Fock ($N<<K/{\cal E}_J $).   
In the Josephson region one expects coherent
superposition of the two BEC's (Schr\"odinger cat states)
to be possible whereas in the Fock region the
two BEC's will be (in some sense) localised
in the two separate wells.
 There is a correspondence between
(\ref{ham}) and the motion of a pendulum \cite{l01}. In the Rabi and Josephson
regimes this motion is semiclassical, unlike the case of the Fock regime.
For both the Fock and Josephson regimes the analogy corresponds to a 
pendulum with fixed length, while in the Rabi regime the length varies.  
An important problem is to study the behaviour in the crossover regimes.

The exact solvability of (\ref{ham}) which we discuss here follows from the
fact that it is mathematically equivalent to the discrete self-trapping
dimer model  
studied by Enol'skii
et al. \cite{esks91}, who solved the model through the algebraic Bethe
ansatz.
The Lax operator appearing in the Yang-Baxter equation 
which underlies  the exact solution 
first appeared in \cite{kt89}.
The consequences of this result are legion. The exact solution 
gives a direct method, through numerical analysis,
to investigate the nature of the system in the crossover regions
between the Fock, Rabi and Josephson regimes.
It also provides a means to test the limits of applicability 
of previous approximate treatments 
\cite{mcww97,sfgs97,zsl98,rsfs99,ji99,mrfss99,ads}. 
Solvability
through the Yang-Baxter equation raises questions about 
the probability distribution of the
energy level spacings \cite{pzbmm}. Moreover, the formulation of the model
through the algebraic version of the Bethe ansatz opens possibilities
for the calculation of time-dependent form factors and correlation functions,
 as achieved in \cite{lz} for $\Delta \mu=0$.  

In this Letter we will show that the exact
solution allows for the analysis of the asymptotic behaviour of the
energy spectrum in the limits of strong and weak tunneling.  
These results are used to extract the asymptotic limits of the quantum
fluctuations of the relative particle number between the two 
BEC's and the degree of coherence in the system for the ground state.
Our approach also makes it  feasible to undertake 
an asymptotic analysis of the system at
finite temperature which gives insight into the relative influence of
thermal and quantum fluctuations (cf. \cite{ps01}).   
We mention that  these findings provide a useful tool in the numerical
evaluation of  the exact solution in the strong and weak tunneling regions.
In contrast to most Bethe ansatz solvable models\cite{kib},
 the model discused
here, as in the case of the Azbel-Hofstadter problem\cite{wz94},
is a quantum mechanical model rather than a quantum field theory; i.e.,
it has a fixed and finite number of degrees of freedom.

{\it Bethe ansatz solution.} 
Following the standard procedure 
of the algebraic Bethe ansatz 
\cite{kib}, we can derive the Bethe ansatz 
equations (BAE). For $N$ total bosons the BAE for (1) can
be written in the form 
$$
\eta^2 (v^2_\a -\omega^2)=
\prod ^N_{\b \neq \a}\frac {v_\a -v_\b - \eta}{v_\a -v_\b +\eta}, 
~~~\a=1,...,N.$$
The parameters $\eta,\,\omega$ which naturally arise in the Bethe ansatz
solution, along with a scaling factor $\K$, 
are related to the coupling constants of (\ref{ham}) through
the 
identification 
$$ K = 2 \K \eta^2, \ \ \
\Delta \mu = -2 \K \eta \omega, \ \ \ 
{\cal {E}_J}= 2 \K.$$
Each set of numbers $\{v_\alpha\}_{\alpha=1}^N$ which is
a solution of the BAE defines an eigenstate of the 
Hamiltonian.
 For details of the derivation of (2) we refer to
\cite{esks91,lz}. For such a solution, the eigenvector has the form 
$$\Psi(v_1,...,v_N)=\prod_{\a=1}^N C(v_\a)\left|0\right>$$ 
with $C(u)=(u-\o+\eta N_2)a_1^{\d}+\eta^{-1}a_2^{\d}$ and 
$\left|0\right>$ the Fock vacuum.   

For each $N$ we expect $N+1$ independent solutions of the BAE. 
Note that in the derivation of the BAE it is assumed that $v_\a$ are
distinct for different $\a$.
This is a result of the Pauli principle for Bethe 
ansatz solvable models as proved by Izergin and Korepin \cite{ik82}
for the one-dimensional $\delta$-function interaction boson gas. 
Their result may be adapted to the present case, which plays an important
role in our asymptotic analysis of the Bethe ansatz solutions. 
{}From the BAE, we may derive the useful identity
\bea
\prod_{\a=1}^m\eta^2(v_\a-\o^2)=\prod_{\a=1}^m\prod_{\b=m+1}^N
\frac{v_\a-v_\b-\eta}{v_\a-v_\b+\eta} \label{id} \eea
which will be used frequently. 

For a given solution to
the Bethe ansatz equations, the corresponding energy 
eigenvalue of the Hamiltonian is 

\bea E&=&-\K\left(\eta^{-2}\prod_{\a}^N(1+\frac{\eta}{v_\a-u})
-\frac{\eta^2N^2}{4} -u\eta N-u^2 \right. \no \\
&&~~~~~~\left.-\eta^{-2}+\o^2+(u^2-\o^2)\prod_{\a}^N(1-\frac{\eta}{v_\a-u})
\right).     \label{nrg} \eea   
Note that this expression is independent of the spectral parameter $u$ which 
can be chosen arbitrarily. The formula simplifies considerably with the 
choice $u=\omega$, by employing (\ref{id}), which yields a polynomial 
form.   
However, for the purpose of an asymptotic analysis 
in the Rabi regime, 
it is more convenient to choose 
$u=0$, 
while for the Fock regime we use $u=\eta^2$.

{\it Asymptotics.} We start our analysis with the Rabi regime where 
$\eta ^2 N <<1$.   
{}From the BAE it is clear that $\eta^2v_{\a}^2\rightarrow 1$ as 
$\eta\rightarrow 0$, so that 
$v_{\a}\approx\pm \eta^{-1}$. 
However, 
when $\eta=0$ we know that the Hamiltonian is diagonalizable 
by using the Bogoliubov transformation, 
from which we can deduce 
that the solution of the BAE corresponding to the ground state 
must have $v_{\a}\approx\eta^{-1}$. 
Excitations correspond to changing the signs of the 
leading terms in the Bethe ansatz roots. To study the asymptotic
behaviour for the $m$th excited state, we set 
\bea v_\a&\approx&-\eta^{-1}+\e_\a+\eta\delta_\a,~~~~~~\a=1,...,m, \no \\ 
v_\a&\approx& \eta^{-1}+\e_\a+\eta\delta_\a,   ~~~~~~~\a=m+1,...,N, 
\label{exp1} \eea 
with the convention that the ground state corresponds to $m=0$.

{}From the leading terms of the BAE for $v_\a,\,\a\leq m$ we find 
\beq
\e_\a=\sum_{\b\neq\a}^{m}\frac{1}{\e_\a-\e_\b}, \label{eps1}
\eeq
which implies
$$\sum_{\a=1}^{m}\e_\a=0,~~~~\sum_{\a=1}^{m}\e_\a^2=\frac{m(m-1)}{2}.
$$
In a similar fashion we have for $m<\a\leq N$ 
\beq
\e_\a=-\sum_{\a\neq\b=m+1}^{N}\frac{1}{\e_\a-\e_\b}, \label{eps2}
\eeq
which implies
$$\sum_{\a=m+1}^{N}\e_\a=0,~~~~\sum_{\a=m+1}^{N}\e_\a^2
=-\frac{(N-m)(N-m-1)}{2}.  $$

It is clear from (\ref{eps1}) and (\ref{eps2})
why the Pauli exclusion principle applies in the 
present case. In the asymptotic expansion for $v_\a$, $\e_\a$ is assumed
finite.  
However, if $v_\a=v_\b$ for some $\a,\,\b$, then $\e_\a=\e_\b$ and 
(\ref{eps1}) and (\ref{eps2})
imply that $\e_\a,\,\e_\b$ are infinite which is a contradiction. Hence 
$v_\a$ must be distinct for different $\a$. Note also that  
for this approximation to
be valid we require $\eta^{-1}>> \e_\a.$ However, we
see that $|\e_\a|$ is of the order of $N^{1/2}$. Thus our approximation
will be valid for $\eta N^{1/2}<<1$, which is precisely the criterion 
for the Rabi region 
and consequently
$N$ cannot be arbitrarily
large for fixed $\eta$, or vice versa. 

Now we go to the next order. 
From (\ref{id})  we find 
\bea
\sum_{\a=1}^m\delta_\a
&=&-\frac{m(m-1)}{4}+\frac{m(m-N)}{2}
-\frac{m\o^2}{2} 
\no \eea
\bea 
\sum_{\a=m+1}^{N}\delta_\a
&=&-\frac{(N-m)(N-m-1)}{4}+\frac{m(m-N)}{2}\no \\
&&~~~~~~
+\frac{(N-m)\o^2}{2}  \no \eea   
which using  (\ref{nrg}) leads us to the result 
$$
\frac{E_m}{\kappa} \ap -N+2m-\frac{\eta^2\o^2(N-2m)}{2}
+\frac{\eta^2N}{4}+\frac{\eta^2}{2}m(N-m).
$$
The energy level spacings $\Delta_m=E_m-E_{m-1}$ are thus 
\bea 
\Delta_m 
&\ap&\kappa\left(2+\eta^2\o^2+\frac{\eta^2}{2}(N-2m+1)\right). \no \eea 
One may check that $\Delta_m /N$ is of the order of $N^{-1}$. 
This indicates that the Rabi regime is semiclassical\cite{l01}.
Our value for the gap between the ground and first excited state 
agrees,
to leading order in $\eta^2 N$,  
 with the Gross-Pitaevskii mean-field theory\cite{sfgs97}
which gives a Josephson plasma frequency of
$\omega_J = 2 \kappa ( 1 + \eta^2 N/2)^{1/2}$.

Now we look at the asymptotic behaviour of the BAE in the Fock regime
$\eta^2 >> N$.  It is necessary to 
distinguish the following cases: (i) $\o=0$ and (ii) $\o \neq
0$.

(i) $\o = 0$. In this case, it is appropriate to consider the permutation
operator $P$ which interchanges the labels 1 and 2 in (\ref{ham}). 
For $\o=0$, $P$ commutes with the Hamiltonian, and any eigenvector of
the Hamiltonian is also an eigenvector of $P$ with
eigenvalue $\pm 1$. Therefore the Hilbert space splits into the direct
sum of two subspaces corresponding to the symmetric and antisymmetric
eigenfunctions. From now on we restrict ourselves to the case when $N$ is
even, i.e., $N=2M$, although a similar calculation is also applicable to 
the case when $N$ 
is odd. A careful analysis leads us to conclude that the ground
state lies in the symmetric subspace. The asymptotic form of
the roots of the BAEs  
for the ground state takes the ``string''-like structure
\bea
v_{j\pm} &\ap& -(M-j) \eta \pm i \frac {C^j_M}{(j-1)!} \eta ^{-(2j-1)} 
\no \\
&&~~~~~~+
M(M+1) \eta ^{-3} \delta _{j1}, 
~~~~~~~~j=1,\cdots, M. \no
\eea
where $C^j_M$ is a binomial coefficient.  For
this asymptotic ansatz to be valid, we require that any term in the
asymptotic expansion should be much smaller than those preceeding. This 
yields $\eta^2 >> N$ which coincides with  the defining condition for the Fock
region. Throughout, the Pauli exclusion principle
has been taken into account to exclude any possible spurious solutions 
of the BAE.  

The above structure  clearly indicates that in the ground state 
the $N$ bosons fuse into
$M$ ``bound'' states and excitations correspond to a breakdown of these 
bound states. Specifically, the first and second excited states
correspond to the breakdown of the bound state at $-(M-1) \eta$, with
the first excited state in the antisymmetric subspace and the second
excited state in the symmetric subspace. Explicitly, we can write down
the spectral parameter configurations for the first two excited states
\bea
 v_{1+} &\ap& -M \eta +a_{1+} \eta ^{-3},~~   
 v_{1-} \ap -(M-1) \eta +a_{1-} \eta ^{-3},\no\\
 v_{j\pm} &\ap& -(M-j) \eta +a_{j\pm} \eta ^{-(2j-1)},
  ~~~~j=2,\cdots,M, \no
\eea
with
\bea
a_{1+} &\ap& -\frac {M+1}{2},~~~~   
a_{1-} \ap \frac {M(M+1)}{2}, \no\\
a_{2\pm} &\ap& \frac {-(M-1)^2 \pm (M-1) \sqrt {13 M^2 +10M +1}}{12}, \no\\
a_{3\pm} &\ap& \pm \frac {(M-1)(M-2) \sqrt {2M(M+1)}}{24}, \no\\
a_{j\pm} &\ap& \frac {M-j+1}{\sqrt {(j+1)j(j-1)(j-2)}} a_{j-1,\pm}, 
~~j=3,\cdots,M, \no
\eea
for the (antisymmetric) first excited state and
\bea
a_{1+} &\ap& -\frac {(M+1)(2M+1)}{2},~~~~ 
a_{1-} \ap -\frac {M(M+1)}{2}, \no\\
a_{2\pm} &\ap& \frac {-(M-1)^2 \pm i (M-1) \sqrt {11 M^2 +14M -1}}{12}, \no\\
a_{3\pm} &\ap& \pm i \frac {(M-1)(M-2) \sqrt {2M(M+1)}}{24}, \no\\
a_{j\pm} &\ap& \frac {M-j+1}{\sqrt {(j+1)j(j-1)(j-2)}} a_{j-1,\pm}, 
~~j=3,\cdots,M, \no
\eea
for the (symmetric) second excited state.
The breakdown of the bound state at $-(M-j)\eta$,  
$j=2,\cdots,M$ results in
the higher excited states.

Substituting these results into (\ref{nrg})
leads us to the asymptotic ground state energy 
$$
E_0 \ap - 2\K \eta ^{-2}M(M+1),
$$
while for the first and second excited states we have
\bea
E_1 &\ap& \K \eta ^2  - \K \eta ^{-2} \frac {M^2 +M -2}{3},\no\\
E_2 &\ap& \K \eta ^2  + \K \eta ^{-2} \frac {5M^2 +5M +2}{3}.\no
\eea
In contrast to the Rabi regime, the Fock regime is not semiclassical, as 
the ratio of the gap $\Delta$ and $N$ is of finite order when
$N$ is large. 

We can perform a similar analysis for odd $N$. In this case, the gap
between the ground and the first excited states is proportional to
$\K \eta ^{-2}$ instead of $\K \eta ^2$. This indicates there
is a strong parity effect in the Fock regime. Its possible physical
implication remains to be explored.

(ii) $\o \neq 0$. In this case the root structure is somewhat more 
complicated than for $\o=0$, so we will not present the details.
We remark however that our 
calculations show that up to order $\eta^{-2}$ the ground 
state energy eigenvalue takes the
same form as in the case $\o =0$. Actually, the leading contribution 
arising from
the $\o$ term appears only as $\o^2 \eta ^{-4}$.
This means that the results presented below are applicable for all
values of $\o$ (or equivalently $\Delta\mu$). 

{\it Quantum fluctuations and coherence factor.} 
Although it is difficult to define rigorously\cite{l01}, 
the relative phase between Bose-Einstein condensates has been useful
in understanding
interference experiments\cite{otfyk,cbfmmtsi,hmwc98}. 
By definition, the relative 
phase $\Phi$ is
conjugate to the relative number of atoms in the two condensates
$n\equiv N_1-N_2$. Using     the Feynmann-Hellman theorem, we find that
$$<\Delta n^2> = 8 \frac {\partial E}{\partial K} -4  ( 
 \frac {\partial E}{\partial \Delta \mu} )^2.$$ 
For the ground state 
in the limit of {\it strong } tunneling (i.e., Rabi regime),
$<\Delta n^2> = N-({\Delta \mu N}/{{\cal E}_J})^2$. 
In the case of {\it weak } tunneling (i.e., Fock regime),
$<\Delta n^2> \ap 2 N(N+2) ( {{\cal E}_J}/{K})^2$. 
The degree of coherence between the two BEC's can be discussed in terms of
\cite{l01} 
$$\a \equiv \frac{1}{2N}
<a_1^\dagger a_2 + a_2^\dagger a_1> = - \frac{1}{N} \frac{\partial E}{\partial
{\cal E}_J }$$ 
where the last identity follows from the
Feynmann-Hellmann theorem.
In terms of  the relative phase $\Phi$, 
$\a \doteq <\cos \Phi >$ \cite{ps01}.
In the strong coupling limit, $\a \ap 1- N^{-1} (\Delta
\mu)^2 / (8 {\cal E}_J)^2$, indicating very close to
full coherence in the ground state. 
In the opposite limit, we have $\a \ap 2(N+2) {\cal
E}_J / K<<1$, indicating the absence of coherence.

The above results give the first order corrections to the results 
presented in 
\cite{otfyk,ps01} for the number
fluctuations and the coherence factor at zero temperature. 
Moreover,  we can extend this approach to gain insight into the
effects of thermal fluctuations. 
This problem was recently studied in \cite{ps01} using Gross-Pitaevskii
theory. 
 In the Fock
regime, the energy gap is large, and so for the low temperature
thermodynamics it is sufficient to consider the first two excited states,
for which the calculations are straightforward. However, in the Rabi
regime the small energy scale requires us to consider the entire
spectrum. The asymptotic form of the partition function is found to be   
(valid for all temperatures) 
\bea
Z&=&\sum_m \exp (- E_m/T) \no \\
&\ap& \frac {\sinh \frac {\b}{2} (N+1)}{\sinh \frac
{\b}{2}}
+ \frac {\b K}{4 {\cal E}_J}  \exp (\b N) {\hat {\cal D}}
\left (\frac {1-\exp \b (N+1)}{1-\exp \b} \right ),
\no \eea
with $\b = -\frac {2\kappa}{T}$ 
(we set Boltzmann's constant $k_B =1$)
and  $\hat {\cal D}$ is the differential operator
\bea
\hat {\cal D} &\equiv& -\frac {(\Delta \mu)^2 N}{4{\cal E}_JK} 
+\frac {N}{4} +\left(\frac{(\Delta\mu)^2}{2{\cal E}_JK}  
+ \frac {N}{2}\right) \frac {\partial}{\partial \b}
-\frac {1}{2} \frac {\partial ^2}{\partial ^2\b} 
. \no \eea
In the usual way it is possible to derive a number of physical
quantities from $Z$. For example, 
the temperature dependent coherence factor is simply 
$$
\a(T) =  \frac{T}{N} \frac {\partial \ln Z}{\partial {\cal E}_J}
$$
and an expression for the particle number fluctuations can be found
in a similar manner.
This provides an analytic means to investigate the thermal effects for
the Rabi regime. In order to investigate the crossover to weaker
tunneling,
it is possible to solve the BAE  numerically to compute these
quantities.  
This will be the subject of future work.

In conclusion, we have shown that the canonical
Hamiltonian describing  quantum tunneling
between a pair of BEC's is integrable in
the sense of the algebraic Bethe ansatz.
An closed form expression was given for all of the energy
eigenstates and
eigenvalues in terms of the solutions of the Bethe ansatz
equations. Our approach makes the calculation
of form factors \cite{lz} and correlation
functions possible without any of the limitations
associated with Gross-Pitaevskii mean-field theory
or the number-phase representation.

We thank J. Corney, G.J. Milburn, and M.D. Gould for helpful discussions. 
This work was supported by the Australian Research Council.
 
\end{multicols}
\end{document}